\documentclass[final,3p,times,twocolumn]{elsarticle}

\usepackage{amssymb}
\usepackage{amsmath}
\usepackage{graphicx}
\usepackage[colorinlistoftodos]{todonotes}
\usepackage{algorithm, xcolor}
\usepackage[font=small]{caption}
\usepackage{url}

\journal{Nuclear Physics A}

\begin{document}

\begin{frontmatter}



\title{Demonstration of Sub-micron UCN Position Resolution using Room-temperature CMOS Sensor}


\author[1]{S. Lin}
\author[1]{J. K. Baldwin}
\author[4]{M. Blatnik}
\author[1]{S. M. Clayton}
\author[10]{C. Cude-Woods}
\author[1]{S. A. Currie}
\author[4]{B. Filippone}
\author[4]{E. M. Fries}
\author[8]{P. Geltenbort}
\author[12]{A. T. Holley}
\author[1]{W. Li}
\author[13]{C.-Y. Liu}
\author[1]{M. Makela}
\author[1]{C. L. Morris}
\author[10]{R. Musedinovic}
\author[1]{C. O`Shaughnessy}
\author[6]{R. W. Pattie Jr.}
\author[7]{D. J. Salvat}
\author[11]{A. Saunders}
\author[9]{E. I. Sharapov}
\author[1]{M. Singh}
\author[4]{X. Sun}
\author[1]{Z. Tang}
\author[1]{W. Uhrich}
\author[4]{W. Wei}
\author[1]{B. Wolfe}
\author[10]{A. R. Young}
\author[1]{Z. Wang \corref{c1}}


\cortext[c1]{Corresponding author. E-mail: zwang@lanl.gov}
\nonumnote{LA-UR-23-25082}
\affiliation[1]{organization={Los Alamos National Laboratory, Los Alamos, NM 87545, USA}}
\affiliation[4]{organization={Kellog Radiation Laboratory, California Institute of Technology, Pasadena, CA 91125, USA}}
\affiliation[6]{organization={East Tennessee State University, Johnson City, TN 37614, USA}}
\affiliation[7]{organization={Center for Exploration of Energy and Matter, Indiana University, Bloomington, IN 47405, USA}}
\affiliation[8]{organization={Institut Laue-Langevin, CS 20156, 38042 Grenoble Cedex 9, France}}
\affiliation[9]{organization={Joint Institute for Nuclear Research, 141980 Dubna, Russia}}
\affiliation[10]{organization={Department of Physics, North Carolina State University, Raleigh, NC 27695, USA}}
\affiliation[11]{organization={Oak Ridge National Laboratory, Oak Ridge, TN 37831, USA}}
\affiliation[12]{organization={Tennessee Technological University, Cookeville, TN 38505, USA}}
\affiliation[13]{organization={Department of Physics, University of Urbana-Champaign, Urbana, IL 161801, USA}}

\begin{abstract}
High spatial resolution of ultracold neutron (UCN) measurement is 
{of growing interest to UCN experiments such as} UCN spectrometers, UCN polarimeters, quantum physics of UCNs, and quantum gravity. Here we 
{utilize physics-informed deep learning to enhance the experimental position resolution} and to demonstrate sub-micron spatial resolutions for UCN position measurements obtained using a room-temperature CMOS sensor, extending our previous work \cite{kuk2021projection, yue2023ultrafast} that demonstrated a position uncertainty of 1.5 microns. We explore the use of the open-source software Allpix Squared to generate experiment-like synthetic hit images with ground-truth position labels. We use physics-informed deep learning by training a fully-connected neural network (FCNN) to learn a mapping from input hit images to output hit position. 
The automated analysis for sub-micron position resolution in UCN detection combined with the fast data rates of current and next generation UCN sources will enable improved precision for 
{future UCN research and applications}.

\end{abstract}




\end{frontmatter}


\section{Introduction}

Ultracold neutrons (UCNs) have a kinetic energy less than 300 neV and are the coldest free neutrons created in the laboratory. The unique properties of UCNs allow them to be a powerful tool in studying the fundamental sciences in nature. UCNs are utilized in several experiments such as determining the precise measurement of neutron lifetime \cite{pattie2018measurement,gonzalez2021improved}, determining the properties of neutron beta decay \cite{Brown2018}, and searching for the neutron electric dipole moment \cite{filippone2018worldwide,pendlebury2015revised,abel2020measurement}. In recent years, there is a growing interest in studying the quantum states of UCN using position sensitive measurements of UCN. The precise measurement of the UCN quantum states have application in the areas of dark energy and dark matter search \cite{jenke2014gravity,pignol2015probing}. 
The study of quantum gravity states also benefits from precise position measurements \cite{dubbers2011neutron, ichikawa2014observation}. For all experiments requiring UCN position-sensitive measurements, a spatial resolution of 1 $\mu$m or less is highly desired.

As UCNs do not directly ionize detectors due to the lack of an electrical charge, their detection depends on a nuclear reaction that generates ionizing charged particles. Position-sensitive measurements of UCN through indirect detection have been previously conducted using ZnS:Ag scintillators and the nuclear reaction with boron-10 ($^{10}$B) \cite{wang2015multilayer,wei2016position,morris2017new}. However, the position resolution is limited by equipment specifications such as the size of the photomultiplier tubes \cite{wang2015multilayer,morris2017new} or the light yield of the imaging camera \cite{wei2016position}. A direct detection scheme using boron-coated solid state detectors is utilized to capture UCN hits in an image or movie format for improved position resolution. 

In this paper, we demonstrate the desired sub-micron position resolution for UCN detection using a room-temperature CMOS sensor. This work improves upon our previous work \cite{kuk2021projection}, which uses a scientific-grade boron-coated (bCCD) sensor, in position resolution as well as the approach in estimating the hit position. In \cite{kuk2021projection}, a 2D Gaussian fit approach is applied to a captured UCN hit image and the hit position is determined to be the centroid. While the performance of the fitting approach resulted in an uncertainty of 1.5 $\mu$m, it does not take into account detector physics. Nonetheless, scientific-grade CCD sensors are very costly and must be operated in low temperatures to suppress dark current. To address this issue, we propose to utilize CMOS sensors due to their lower manufacturing cost as well as their capability of operating at room-temperature. In addition, CMOS sensors are typically manufactured with a smaller pixel size compared to CCD sensors, which naturally improves spatial resolution. However, the true hit position data from captured images will not be available for any chosen detector. To address the lack of ground-truth data, we leverage a silicon detector simulation framework Allpix Squared \cite{spannagel2018allpix}, which can generate synthetic data containing Monte Carlo ground-truth information including hit position. To address the issue of detector physics during hit position estimation, we use deep learning to model the underlying physics by training a fully connected neural network (FCNN) to learn a mapping from input images to output ground-truth labels. Our trained 
FCNN model enjoys predicting sub-pixel and sub-micron position resolution.

The rest of this paper is summarized as follows. In Section 2, we introduce the direct UCN detection method using solid state detectors and the nuclear reaction with $^{10}$B. To address the lack of position data, we discuss generating synthetic data with ground-truth position information using the simulation framework Allpix Squared. To learn a underlying detector physics, we discuss the use of deep learning to learn a mapping from input images to ground-truth labels. In Section 3, numerical results demonstrate that the deep learning model achieves sub-pixel and sub-micron position resolution. Lastly, Section 4 concludes this paper.


\begin{figure}[t]
\centering
\includegraphics[width=0.75\linewidth]{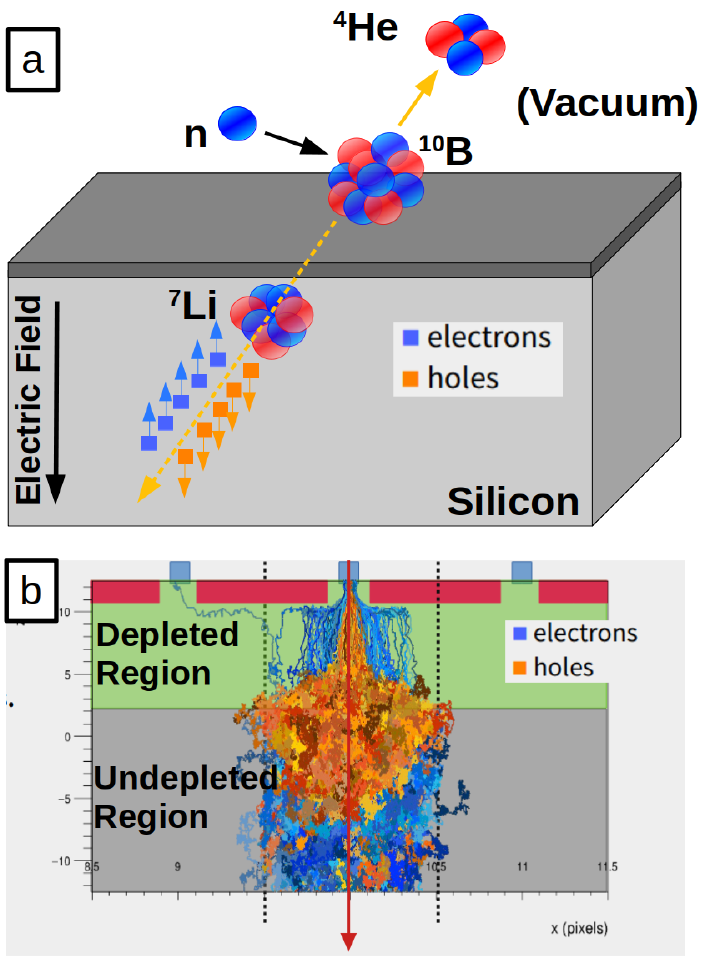}
\caption{(a) Direct UCN detection scheme using a silicon detector and the nuclear reaction with $^{10}$B, which produces two charged particles $\alpha$ and $^7$Li. One charged particle is captured and stopped by the detector to produce e-h pairs along its trajectory. The charge carriers propagate through the silicon detector with direction dependent on the internal electric field direction. (b) An example MAPS sensor simulated using Allpix Squared by Allpix Squared Authors \cite{spannagel2022pixel}. In the depleted region (green), charge carriers drift in the direction of the defined electric field, while thermal diffusion is dominant in the undepleted region (gray).}
\label{fig:ucn_detection_principle}
\end{figure}

\section{Methods}
In this section, we describe the the experimental UCN capture scheme using a solid state silicon detector and the nuclear reaction with $^{10}$B. To attain sub-micron position resolution, we first use the open-source silicon detector simulation software Allpix Squared for data generation. Next, we leverage deep learning for UCN hit prediction by training a FCNN to learn a mapping from input hit images to ground-truth labels. 

\subsection{Solid State Detector using $^{10}\text{B}$}
The working principle of the direct UCN detection scheme using a thin $^{10}$B film deposited on a silicon detector is shown in Figure \ref{fig:ucn_detection_principle}a. Due to the low kinetic energy of UCNs, their detection requires a nuclear reaction with a conversion layer to generate high energy charged particles \cite{pietropaolo2020neutron}. The proposed detection scheme uses a isotopically purified $^{10}$B film up to 100 nm thick as the conversion layer. The $^{10}$B film is deposited on the silicon detector using the electron-evaporation deposition process described in \cite{wang2015multilayer}. The resulting neutron capture reaction generates high energy $\alpha$, $^7$Li, and $\gamma$-ray particles as follows
\begin{equation*}
	\begin{aligned}
		n + ^{10}\text{B} &\rightarrow ^4 \text{He} (1.48 MeV) + ^7 \text{Li} (0.84MeV) \\
		&~~~~~~~+ \gamma (0.48 MeV) ~~~&(94\%) \\
		n + ^{10}\text{B} &\rightarrow ^4 \text{He} (1.78 MeV)+ ^7 \text{Li} (1.02 MeV) ~~~&(6\%)
	\end{aligned}
\end{equation*}
The detector captures the UCN hit when the $\alpha$, $^7$Li, and $\gamma$-ray particles are stopped within the active silicon layer and generate charge carries, also known as electron-hole (e-h) pairs, along its trajectory. The charge creation threshold of silicon is approximately 3.6 eV per e-h pair. Influenced by the internal electric field, one charge carrier type will propagate through the active silicon layer to be collected at the potential wells. Electrons will travel in the direction opposite of the electric field direction, while holes will travel in the same direction. The collected charges are converted to a digitized value and are read out by the detector to generate the UCN hit image. 

To better visualize the charge carrier movement within the silicon detector, Figure \ref{fig:ucn_detection_principle}b illustrates a simulation of the charge carrier propagation within a monolithic active pixel sensor (MAPS) by the authors of Allpix Squared \cite{spannagel2022pixel}. The depleted region (green) represents the detector sensitive volume, where the internal electric field is present and charge carrier drift is dominant. The electrons drift towards and are collected at the collection diodes (blue squares), while the holes drift towards the undepleted region. All charges generated within the depleted region will be collected at the collection diodes such that the charge collection efficiency is maximum. The undepleted region is the electric field free region of the detector and thus, charge transit is dominated though a thermal diffusion process. Hence, the resulting charge cloud is larger as the charge carriers spreads to neighboring pixels. However, charge carriers in the undepleted region may not diffuse back into the depleted region to be collected at the collection diodes and may be subject to recombination losses. Therefore, the charge collection efficiency is reduced in the undepleted region.


\begin{figure}
\centering
\includegraphics[width=0.95\linewidth]{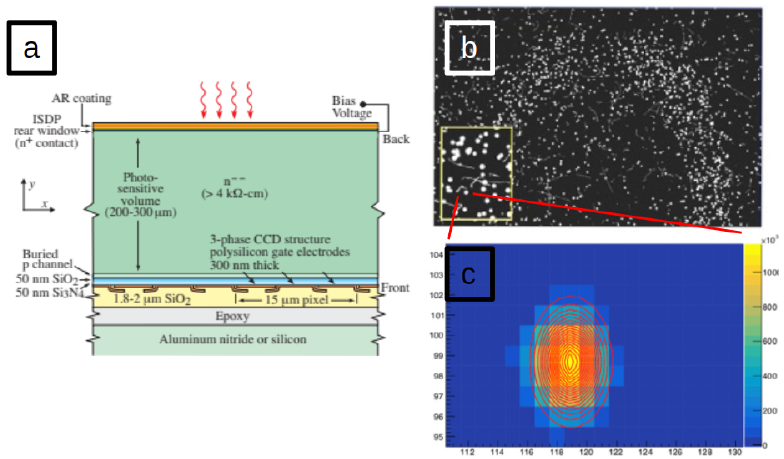}
\caption{(a) Cross-section of the bCCD built by LBNL. (b) A raw bCCD image showing captured UCN hits with a yellow box showing a zoomed in section of the image. (c) A zoomed in view of a single UCN hit. The plotted red contours show a 2D Gaussian fit to the hit.}
\label{fig:bccd}
\end{figure}

\subsection{Detection using bCCDs}

In our previous work \cite{kuk2021projection}, a scientific grade boron-coated CCD (bCCD) was used to detect UCNs. The bCCD used in the experiment is similar to the CCDs used in the Dark Energy Camera (DECam) project \cite{flaugher2015dark}. The sensors were built by the Lawrence Berkeley National Laboratory (LBNL) \cite{holland2003fully} and has been extensively characterized by Fermilab for the DECam project \cite{estrada2010focal}. The bCCD sensor is a fully depleted, back-illuminated sensor fabricated from high-resistivity silicon with a thickness of 250 $\mu$m. The sensor has 8 million (2k $\times$ 4k) pixels with a pixel pitch of $15 \times 15$ $\mu$m$^2$. Fabricated from $10k\Omega$ resistivity silicon, a bias voltage of 25 V allows for fully depleted operation. The sensor is cooled to 140 K to suppress dark current. Figure \ref{fig:bccd}a shows a cross-section of the bCCD sensor. The incoming UCNs hit the sensor back-side with the $^{10}$B coating. The generated charged particle, $\alpha$ or $^7$Li, penetrates and interacts with the silicon layer to create e-h pairs. Under the effect of the internal electric field, the holes created near the sensor back-side travel through the full silicon thickness to be collect at the potential wells at the sensor front-end. Figure \ref{fig:bccd}b shows a hit image captured by the bCCD, while Figure \ref{fig:bccd}c shows a close up of a single hit fitted with a 2D Gaussian fit. The estimated centroid of the 2D Gaussian fit for identified UCN hits was used to estimate the hit position with an uncertainty of 1.5 $\mu$m.

\subsection{Detection using CMOS}

\begin{figure}
\centering
\includegraphics[width=0.95\linewidth]{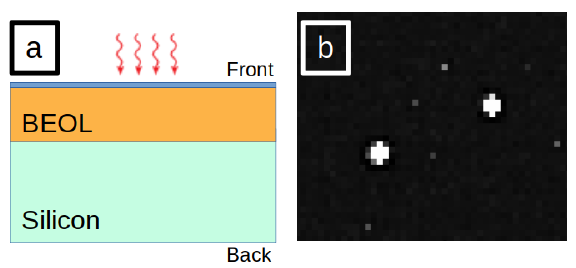}
\caption{(a) Simple cross-section of the DMM CMOS sensor. Note the dimensions are not to scale. (b) Sample of a raw DMM CMOS image that shows two captured UCN hits.}
\label{fig:cmos}
\end{figure}

While bCCDs can be used to detect UCNs, these scientific-grade sensors are very costly. A more cost efficient solution is to utilize off-the-self CMOS sensors for UCN detection. In addition to the cost benefits, CMOS sensors can be operated at room temperature (293 K) and have a smaller pixel pitch. In this experiment, we use the DMM 27UJ003-ML CMOS camera by the Imaging Source. Different from the bCCD, the CMOS sensor is front-illuminated, is not fully depleted, and has a pixel pitch of $1.67 \times 1.67$ $\mu$m$^2$. Furthermore, the sensor has not been extensively characterized and thus, known sensor parameters are extracted from the datasheet. Figure \ref{fig:cmos}a shows an assumed simple cross-section of the CMOS sensor. The incoming UCNs hit the sensor front-end with the $^{10}$B coating. The generated charged particle, $\alpha$ or $^7$Li, needs to travel through the back-end-of-line (BEOL) or metal layer in order to reach the active silicon layer. The BEOL layer thickness varies between different CMOS manufacturers. A thick BEOL layer will fully stop the lower energy $^7$Li particles and thus, only $\alpha$ particles will reach the active silicon layer. Figure \ref{fig:cmos}b shows two experimental hits captured by the DMM camera.

Similar to the bCCD experiment, a 2D Gaussian fit can be fitted to each identified UCN hit to estimate the hit position. However, this fitting approach does not take into account detector physics. Meanwhile, the ground-truth hit position is not available. We propose to overcome this challenge by synthetically generating UCN hits and their corresponding ground-truth hit position by using the silicon detector framework Allpix Squared discussed next.

\begin{figure}[t]
\centering
\includegraphics[width=0.97\linewidth]{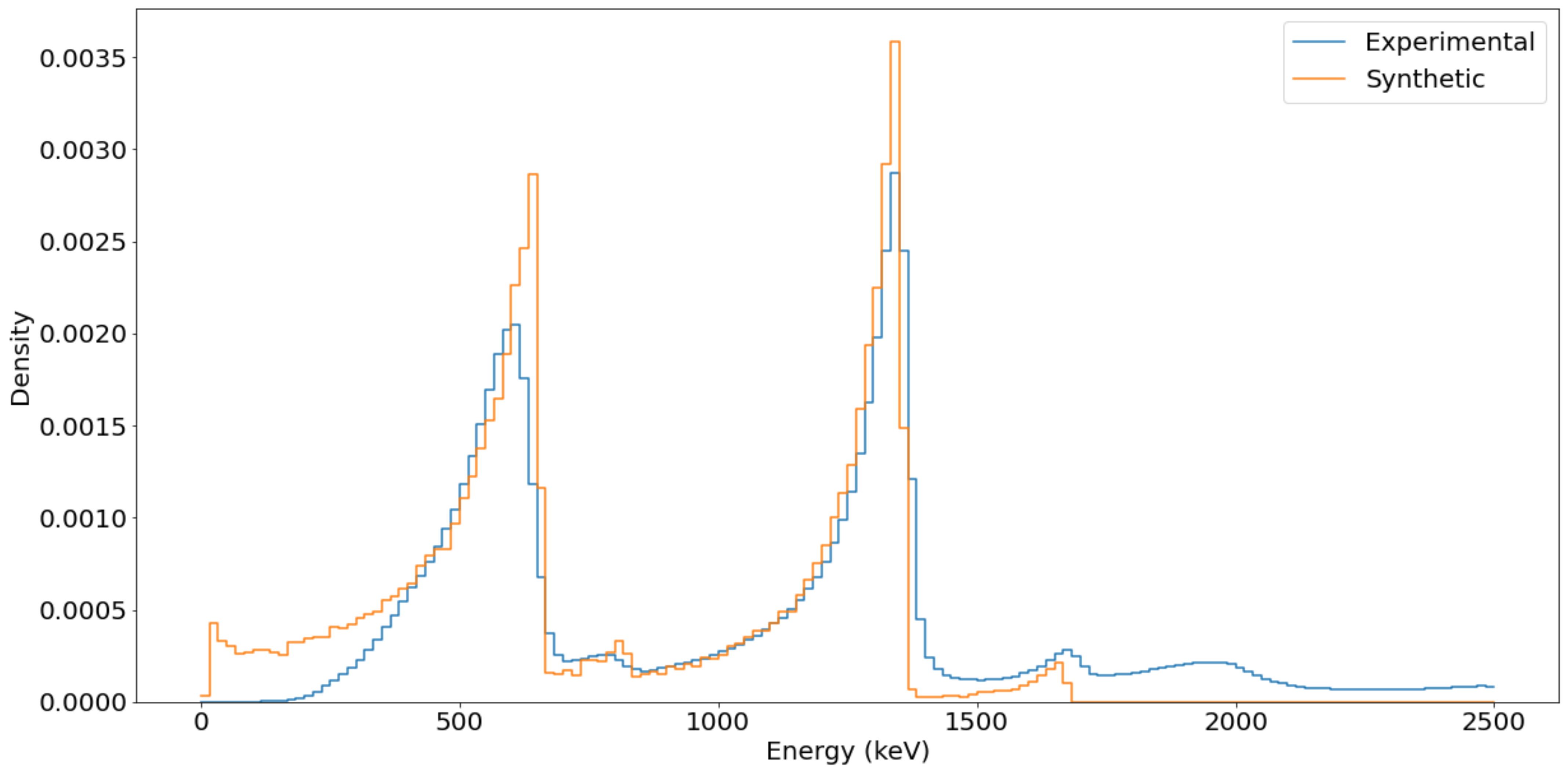}
\caption{Comparison of the captured UCN energy spectrum by the experimental bCCD images (blue) and the Allpix Squared synthetic images (orange). The peaks shown correspond to $^7$Li = 0.84 MeV and $\alpha$ = 1.47 MeV.}
\label{fig:bccd_spec}
\end{figure}

\begin{figure}[tbh!]
\centering
\includegraphics[width=0.8\linewidth]{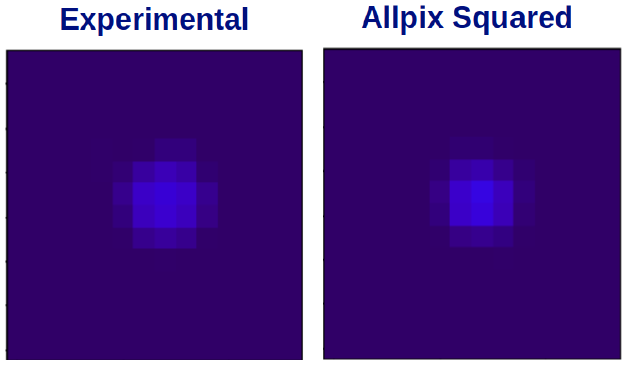}
\caption{Comparison between (left) an experimental bCCD image and (right) the best matched synthetic image generated by Allpix Squared.}
\label{fig:bccd_hits}
\end{figure}

\subsection{Data Generation with Allpix Squared}

Allpix Squared is an open-source silicon detector simulation tool that implements end-to-end Monte Carlo simulations from incident particle detection to digitized detector output \cite{spannagel2018allpix}. The simulation framework is comprised of several stages and different algorithms for detector physics such as charge transport and front-end electronics modeling. In addition, the framework is built upon Geant4 \cite{agostinelli2003geant4} to simulate the interaction of the incident particle with the sensor materials. At each stage of the simulation process, Allpix Squared stores the Monte Carlo truth information such that the entire simulation history can be traced from the initial particle to output image. In Section \ref{sec:dl}, this ground-truth information is extracted and exploited for training a FCNN to learn the underlying detector physics by learning a mapping from input hit images to output hit position. 

To demonstrate the silicon detector modeling capability of Allpix Squared, we model the fully characterized bCCD sensor and compare the synthetic and experimental data. Figure \ref{fig:bccd_spec} plots and compares the captured energy spectrum of the experimental hit images (blue) and the synthetically generated ones (orange). The spectrums were fitted to align the peaks of the 1.48 MeV $\alpha$ particle. Clearly, the synthetic spectrum well captures the experimental. Meanwhile, Figure \ref{fig:bccd_hits} compares the experimental and synthetic hit images and demonstrates that the simulation framework can generate synthetic hits that are similar to the experimental. Nonetheless, we show that a fully characterized detector can be accurately modeled in Allpix Squared.

\begin{figure}
\centering
\includegraphics[width=0.85\linewidth]{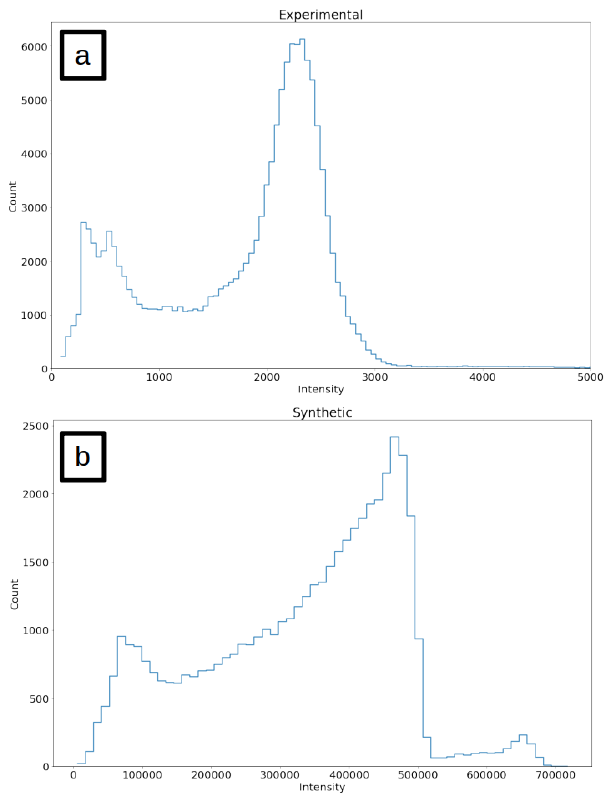}
\caption{(a) The raw experimental spectrum captured by the DMM CMOS sensor and (b) the raw synthetic spectrum captured by Allpix Squared. The x-axis values denotes the raw intensity values for each plot.}
\label{fig:cmos_spec}
\end{figure}

Different from the bCCD, the CMOS sensor is not fully characterized. The few known detector parameters are extracted from the datasheet, while unknown detector parameters need to be tuned. Figure \ref{fig:cmos_spec}a plots the experimental energy spectrum and Figure \ref{fig:cmos_spec}b the synthetic one. The current tuned values for the unknown detector parameters results in a synthetic spectrum with a trend that is similar to the experimental. A more accurate CMOS model can be obtained with further parameter tuning.

\begin{figure}[thb!]
\centering
\includegraphics[width=0.85\linewidth]{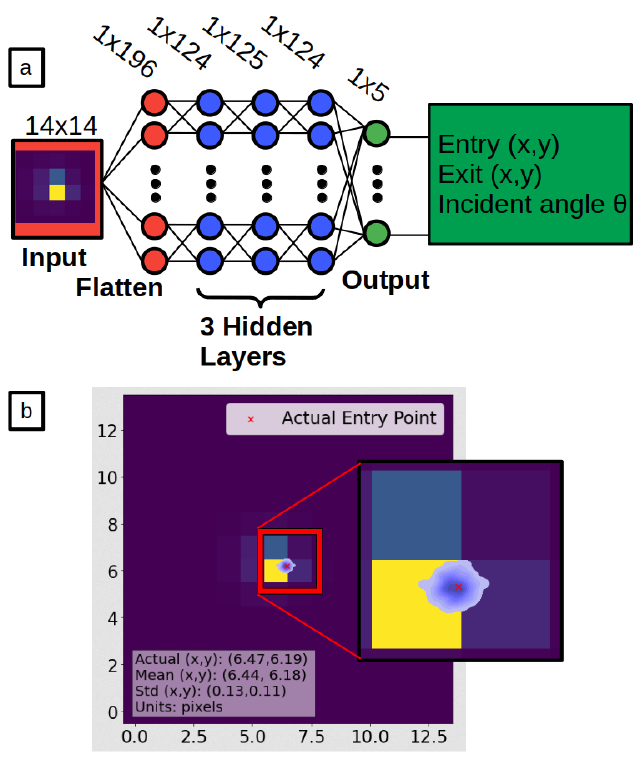}
\caption{(a) Overview of the FCNN architecture with 3 hidden layers. The network takes as input UCN hit images of size $14 \times 14$ pixels and outputs the entry and exit $(x,y)$ coordinates as well as the incident angle $\theta$. (b) An example entry point prediction for an input image. The red `x' denotes the actual (ground-truth) entry point and the blue KDE plot shows the FCNN prediction with dropout layers enabled. The prediction for this example clearly obtains sub-pixel resolution.}
\label{fig:fcnn}
\end{figure}

\subsection{Deep Learning for Position Resolution} \label{sec:dl}

Allpix Squared can be used to generate large datasets of UCN hit images and their corresponding ground-truth labels such as hit position. We propose to model the underlying complex detector physics by using deep learning. A sub-field of machine learning and artificial intelligence, deep learning is widely used to learn underlying structures and patterns in high-dimensional data by leveraging neural networks for their superior nonlinear approximation capability \cite{lecun2015deep}. Deep learning has been popularly applied in fields such as image and speech recognition \cite{guo2016deep, nassif2019speech}, robotics  \cite{pierson2017deep, fujiyoshi2019deep}, and medical imaging \cite{lee2017deep}. We apply deep learning to model the detector physics by learning a mapping from input UCN hit images to output ground-truth labels. Specifically, we train a FCNN with three hidden layers as shown in Figure \ref{fig:fcnn}a. The input UCN hit images are of size $14 \times 14$ pixels that are flattened into a 1D vector of size $1\times 196$. Following the flattened layer, there are 3 hidden layers with 124, 125, and 124 neurons, respectively. Lastly, the output layer consists of the ground-truth labels consisting of the entry and exit coordinate as well as the incident angle $\theta$, the angle at which the particle enters the detector. Not shown in the FCNN architecture, dropout layers are implemented after each hidden layer to mitigate model overfitting issues and to model uncertainty quantification during testing.

\section{Results and Discussion}

We use Allpix Squared to generate a large dataset for the CMOS detector. The dataset consists of approximately 50,000 UCN hit images and their corresponding ground-truth labels, consisting of the entry and exit coordinates as well as the incident angle $\theta$. To train and test the neutral network in Figure \ref{fig:fcnn}a, the dataset is split into 60\% for training, 20\% for validation, and 20\% for testing. We use the Pytorch library \cite{paszke2019pytorch} to train the FCNN to learn a predictive mapping between the input images and output labels using the training and validation datasets. The trained model is then validated using the test dataset by making predictions on the test images. During the testing phase, the dropout layers in the FCNN are enabled to allow for uncertainty quantification for label prediction. For each test image, we feed the image into the trained FCNN for 500 times to simulate Monte Carlo runs. Figure \ref{fig:fcnn}b shows an example FCNN prediction as a kernel density estimate (KDE) on the entry point position (in blue) for one synthetic hit image. Clearly, the FCNN prediction obtains sub-pixel position resolution. Furthermore, it obtains \textit{sub-micron} position resolution as the CMOS pixel size is $1.67$ $\mu$m.

For each image in the test dataset, we compute the mean and standard deviation of the predicted labels and compare the predicted values to the ground-truth. For each label, we compute the average absolute deviation of the predicted mean value and the true value as well as the average predicted standard deviation. Table \ref{table:1} summarizes the overall FCNN model performance on the test dataset. The model achieves a maximum mean deviation of 0.18 pixels (0.3 $\mu$m) and standard deviation of 0.12 pixels (0.2 $\mu$m) for predicting the entry and exit point position. Thus, the FCNN achieves both sub-pixel and sub-micron position resolution.

\begin{table}
\centering
\caption{Summary of FCNN performance on the test dataset. The second column tabulates the average absolute deviation of each label prediction where $N$ denotes the total number of test samples, $\hat{x}_n$ denotes the predicted mean for sample $n$, and $x_n$ the true value. The third column tabulates the average predicted standard deviation where $\hat{\sigma}_n$ denotes the predicted standard deviation for sample $n$.}
	\begin{tabular}{| c | c  | c |}
	\hline
	\textbf{Output Label} & \textbf{$\frac{1}{N} \sum_N |\hat x_n - x_n|$} & \textbf{$\frac{1}{N} \sum_N \hat \sigma_n$}\\
	\hline
	Entry x (pixels) & 0.13 & 0.12 \\
	Entry y (pixels) & 0.13 & 0.11 \\
	Exit x (pixels)& 0.17 & 0.11\\
	Exit y (pixels)& 0.18 & 0.10 \\
	$\theta$ (degrees) & 4.48 & 3.21 \\
	\hline
	\end{tabular}
	\label{table:1}
\end{table}

\section{Conclusion}

A direct UCN detection scheme using solid state detectors and a thin $^{10}$B film was described. The detection scheme was applied to scientific-grade bCCD sensors as well as the DMM CMOS sensor to obtain UCN hit images. Due to the lack of ground-truth position information, Allpix Squared is used to generate synthetic data consisting of images and corresponding ground-truth labels by modeling the silicon detectors as accurately as possible. A FCNN model is trained to learn underlying detector physics by learning a mapping from input hit images to output ground-truth labels. The trained FCNN achieved both sub-pixel and sub-micron resolution for UCN position resolution.

\section*{Declaration of competing interest}

The authors declare that they have no known competing financial interests or personal relationships that could have appeared to influence the work reported in this paper.

\section*{Acknowledgements}

SL and ZW wish to thank Dr. Don Groom and Dr. Steve Holland, both from Lawrence Berkeley National Laboratory, for bCCD simulating discussions. SL and ZW wish to thank Dr. Simon Spannagel for help with Allpix Squared. 

{This work is supported in part by the LANL LDRD program.} 



\bibliographystyle{elsarticle-num} 
\bibliography{ref}


%
%
%
\end{document}